\documentclass[11pt]{article}

\usepackage{latexsym}
\usepackage{graphicx}
\usepackage{color}

\begin{document}
\title{A new instability for finite Prandtl number rotating
  convection with free-slip boundary conditions}
\author {Y. Ponty, T. Passot and P.L Sulem\\
CNRS URA 1362, Observatoire de la C\^ote d'Azur\\
B.P. 229, 06304 Nice Cedex 04, France}
\date{}
\maketitle

\centerline{ \it Phys. Fluid {\bf 9}, 67 (1997)}

\begin{abstract}
Rolls in finite Prandtl number rotating convection with free-slip
top and bottom boundary conditions are shown 
to be  unstable with respect to small angle perturbations
for any value of the rotation rate. This
instability is driven by the horizontal 
mean flow whose estimation requires a special singular perturbation
analysis.

\medskip
\end{abstract}

 \setcounter{equation}{0}
\section{Introduction}
Rayleigh-B\'enard convection in a plane layer heated from below and
rotating about a vertical axis, has been the object of special 
attention motivated by both astrophysical and geophysical applications, 
and by the existence of additional instabilities occuring in this system.
In the case of free-slip top and bottom boundary conditions,
K\"uppers and Lortz 
\cite{KL} showed, using by a perturbation analysis near threshold, 
that when in an infinite Prandtl number fluid, the Taylor number
(which measures the rotation rate) exceeds the 
critical value 2285, two-dimensional rolls are unstable  with respect
to perturbations of the form of a similar pattern rotated by 
an angle close to $58^{o}$. 
This instability which is also present with no-slip
boundaries \cite{Clever}, leads in the case of extended
systems to the formation of chaotically evolving patches of parallel
rolls  \cite{Cross1}, \cite{Ning}, \cite{Cross2},\cite{Hu},\cite{Busse}. 

Convection at moderate Prandtl number with no-slip top and bottom
boundary conditions, was addressed in \cite{Kuppers}, \cite{Clever} and
\cite{Knobloch}, and the K\"uppers-Lortz instability was shown to
occur at a critical Taylor number lower than in the infinite 
Prandtl number limit. 
Free-slip boundaries were considered by
Swift (cited in \cite{Knobloch}) 
who noted that the usual perturbative calculation of the
growth rate leads to a divergence in the limit of perturbations
quasi-parallel to the basic rolls.
The present paper is mostly concerned with a revisited analysis of this
problem, leading to a uniformly valid expression of the instability
growth rate.  We show in particular that for any finite Prandtl
number and rotation rate, straight parallel rolls are unstable when
the angle associated to the perturbation is small enough.

In Section 2, steady convective rolls in a rotating frame are
constructed perturbatively near threshold. Section 3 is devoted to
the computation of the instability growth rate for finite angle
perturbation, an analysis which, at finite Prandtl number, 
breaks down in the small angle limit. 
In Section 4, we present a special analysis in 
the resulting small angle ``boundary layer'',
where the interaction of the basic rolls with quasi-parallel perturbations
leads to almost space-independent contributions which
become resonant in the zero angle limit. These terms are removed by
prescribing a quasi-solvability condition the
marginal mode of quasi-constant horizontal velocity. A uniform expression
for the instability growth rate 
is then derived and a new ``small-angle instability'' is obtained. 
The sensitivity of the instability growth rate to the
Prandtl and Taylor numbers is analyzed. 
Qualitative features of this instability and its
nonlinear development are briefly described in Section 5.

\setcounter{equation}{0}
\section{Steady convective rolls in a rotating frame}

The Boussinesq equations in a horizontal fluid layer heated from
below and  rotating around a vertical axis  
${ \hat  {\bf z}}$, are written in the non-dimensional form
\begin{eqnarray}
&&\Delta {\bf  u}  +  { \hat  {\bf z}}  \vartheta - \nabla \Gamma -
\tau  { \hat  {\bf z}} \times  {\bf  u}    =   
  {P_r}^{-1}  ( {\bf  u}. \nabla {\bf  u}  +
  \frac{\partial}{\partial t} 
  {\bf  u}  ) \label{eq:bb1}\\
&&  \nabla .{\bf  u}   =  0 \label{eq:bb2} \\
&& \Delta \vartheta  + R_a   { \hat  {\bf z}} .  {\bf  u} 
  = {\bf u}.\nabla \vartheta  + \frac{\partial}{\partial t}
 \vartheta ,
 \label{eq:bb3} 
\end{eqnarray}
\noindent
where the vertical diffusion time is taken as
time unit. We assume a Prandtl number $P_r > 0.6766 $, to prevent
over-stability \cite{Chand}. The other parameters are  the Rayleigh
number $R_a$ and the square root $\tau$ of the Taylor number 
(equal to twice the Rossby number) which, to be specific, 
is taken positive (anti-clockwise rotation).

Proceeding as in \cite{KL}, we introduce the operators $\Lambda  =
\nabla \times ( \nabla \times . ) $ 
 and  $ \Upsilon =  \nabla \times .$, and express the velocity 
${\bf u} = ( u, v, w)^t $  in terms of two scalar fields
 $\phi$ and $\psi$, in the form
${\bf  u} =  \Lambda (\phi {\hat {\bf z}}) + \Upsilon
(\psi {\hat{\bf z}}) = ( \partial_z \partial_x \phi  + \partial_y
\psi,\partial_z \partial_y \phi  - \partial_x \psi, - \Delta_h
\phi)^t$, where $\Delta_h = \partial_{xx} + \partial_{yy} $.
Applying  the operators ${\hat{\bf z}}.\Lambda$ and ${\hat{\bf
    z}}.\Upsilon$ on eqs. (\ref{eq:bb1})-(\ref{eq:bb3}), we obtain
\begin{equation}  (U + R_a G)  X = Q(X,X) +
  \frac{\partial}{\partial t}  V X,  \label{eq:mb1}
\end{equation}
with
  \begin{eqnarray}
X = \left [ \begin{array}{c}\phi \\  \psi \\ \vartheta  \end{array}
\right ] , \    
Q(X,X') =\left [ \begin{array}{c}  {P_r}^{-1} {\hat {\bf z}}.\Lambda
    ({\bf u}\cdot\nabla {\bf u'}) \\ 
- {P_r}^{-1}  {\hat {\bf z}}. \Upsilon  ( {\bf u}\cdot\nabla {\bf
  u'})\\  {\bf u}\cdot \nabla  \vartheta'   
\end{array} \right ] , \  
  G= \left [ \begin{array}{ccc}   0 & 0 & 0 \\ 0 & 0 & 0 \\
      -\Delta_h & 0  & 0 \end{array} \right ] ,   
\nonumber
\end{eqnarray}
\begin{eqnarray}
 U = \left [ \begin{array}{ccc}   \Delta^2 \Delta_h  & - \tau
     \partial_z \Delta_h & - \Delta_h  \\
 \tau \partial_z \Delta_h  & \Delta \Delta_h & 0  \\
  0 & 0 & \Delta \end{array} \right ] ,\ 
  V= \left [ \begin{array}{ccc} {P_r}^{-1} \Delta \Delta_h  & 0  &
      0  \\ 0 &  {P_r}^{-1} \Delta_h & 0 \\  0 & 0 & 1 \end{array}
  \right ]  \nonumber . 
\end{eqnarray}
For free-slip boundary conditions, 
$\vartheta=\phi = \partial_{zz} \phi = \partial_z \psi  = 0$ 
in the planes $ z = \pm \frac{1}{2}$.

A stationary solution of eq. (\ref{eq:mb1}) is computed
perturbatively near the convection threshold  by expanding 
$ R_a  =R_0 + \epsilon R_1 + \epsilon^2 R_2 + \cdots \label{eq:R}$
and $ X  = \epsilon X_1 + \epsilon^2 X_2 + \epsilon^3 X_3 + \cdots
.$ or, more explicitly, when taking into account the boundary 
conditions satisfied by the individual components,
\begin{eqnarray}
&&\phi= \epsilon \phi_1 \cos \pi z + \epsilon^2 \phi_2 \sin 2 \pi z +
\cdots \\ 
&&\psi= \epsilon \psi_1 \sin \pi z  +\epsilon^2 (\psi_0 + \psi_2
\cos 2 \pi z  ) + \cdots \\
&&\vartheta= \epsilon \vartheta_1 \cos \pi z  +
 \epsilon^2 \vartheta_2 \sin 2 \pi  z  + \cdots  .
\end{eqnarray}  
Introducing the linear operator  $ L = U + R_0 G$, we get at the
successive orders of the expansion, 
 \begin{eqnarray}
        && L X_1  =  0                                \label{eq:dv1}
        \\  
        && L X_2  =  -R_1 G X_1 + Q(X_1,X_1)
        \label{eq:dv2} \\ 
        && L X_3  =  -(R_1 G X_2+ R_2 G X_1) + Q(X_1,X_2) +
        Q(X_2,X_1)  \label{eq:dv3}. 
 \end{eqnarray}

For a solution in the form of two-dimensional rolls with a critical
wavenumber $| {\vec k_1} | = k$, given by the real solution of
\begin{equation}
2(\frac{k^2}{\pi^2})^3+3(\frac{k^2}{\pi^2})^2=1+\frac{\tau^2}{\pi^4}, 
\end{equation}
the critical Rayleigh
number is $R_0 = \frac{(k^2+\pi^2)^3+\tau^2 \pi^2}{k^2}$ \cite{Chand}.
To simplify the writing, we denote by  
\begin{equation}
Z(\alpha,\beta,\gamma) = 
( \alpha \cos \pi z, \beta \sin\pi z, \gamma \cos \pi z )^t , 
\end{equation}
vectors corresponding to fundamental modes in the vertical
direction and obeying the boundary conditions prescribed on $X$.
An element of the null space of $L$ is then given by
\begin{equation}
v({\vec k}) = Z(c_1,c_2,c_3) e^{i {\vec k}.{\vec x}}  ,
\end{equation}
with
 $ c_1 = 1,\ \  c_2 = - \frac{\tau \pi}{k_p^2},\ \  c_3 =\frac{R_0
   k^2}{k_p^2}, $ and $   k_p^2 = k^2 + \pi^2 =
 \sqrt{\frac{R_0}{3}} \ , 
$ 
and the leading order solution reads
\begin{equation}
 X_1 = A v({\vec k_1}) + c.c. ,
\end{equation} 
where the amplitude A will be determined by a solvability condition
arising at a higher order. For this purpose, it is convenient to 
introduce the inner product  
\begin{equation}
\langle X, X' \rangle = R_0   \int \phi^* \phi' d{\vec x} +
 R_0   \int \psi^* \psi' d{\vec x} 
+ \int \vartheta^* \vartheta'  d{\vec x},
\end{equation}
for which the operator $L$ is self-adjoint.

Using the notation 
\begin{equation}
Q(X_i,X_j) + Q(X_j,X_i)  = (1 + \delta_{ij}) (Q^{(1)}_{i,j},
Q^{(2)}_{i,j},  Q^{(3)}_{i,j})^{t}  ,
\label{eq:notation}
\end{equation}
we have in eq. (\ref{eq:dv2}),
\begin{equation}
Q(X_1,X_1)=\left [ \begin{array}{c}  Q^{(1)}_{1,1} \\ Q^{(2)}_{1,1} 
\\ Q^{(3)}_{1,1}  \end{array}  \right ]
=\left [ \begin{array}{c}  0 \\
2 \frac{\tau}{P_r} \pi^2 \frac{k^4}{k_p^2} ( A^2 e^{2
  i {\vec k_1}. {\vec x}} + c.c.)  \\
- 2 R_0  \pi \frac{k^4}{k_p^2} \sin 2 \pi z |A|^2
\end{array} \right ].
\end{equation}
The solvability condition for eq. (\ref{eq:dv2}) (obtained by taking
the inner product of this equation with $v(\vec k_1)$), requires  $R_1 =0 $.  
Defining the operator  $P = \Delta^3 +\tau^2 \partial_{zz} - R_0
\Delta_h $,  eq. (\ref{eq:dv2}) reduces to
\begin{equation} 
 \Delta_h P \phi_2 =
 \Delta Q^{(1)}_{1,1} + \tau\partial_{z} Q^{(2)}_{1,1} + \Delta_h
 Q^{(3)}_{1,1}    .
 \label{eq:phi2}
\end{equation}
The right hand side vanishing identically, we get
$ \Delta_h\phi_2 = 0$, since elements of the null space of $P$,
already included in $\phi_1$, are not needed in $\phi_2$.

For the two other components of $X_2$, one easily checks that
$\psi_2 = 0$, $\psi_0 = \Psi_0  e^{2 i {\vec k_1}.{\vec x}} + c.c. $
and  $\vartheta_2 = \Theta_2  $ with
$ \Psi_0 = 
 \frac{\tau \pi^2 }{8 P_r k_p^2} A^2$  and
$\Theta_2 = 
 \frac{R_0 k^4 }{2 \pi  k_p^2} |A|^2 $.
This enable us to compute
$$  Q(X_1, X_2) + Q(X_2, X_1) =  \left (
\begin{array}{l} 0 \\ \frac{4  k^4 \pi}{P_r} 
\Psi_0 ( A^* e^{ i   {\vec k_1}. {\vec x}} -
3 A e^{3 i  {\vec k_1}. {\vec x}} ) \sin \pi z + c.c.  \\
2  \pi k^2 \Theta_2 A    
e^{i {\vec k_1}.{\vec x}} \cos \pi z  \cos 2 \pi z  + c.c.  
\end{array} \right ). $$
The solvability condition of eq. (\ref{eq:dv3}) then reduces to 
$ R_2 = r_2 |A|^2$ or equivalently, 
$\epsilon^2|A|^2 = \frac{R_a -R_0}{r_2}$ , with $r_2 
= \frac{ 1}{2 k_p^2} (R_0 k^4 -\frac{\tau^2 \pi^4}{{P_r}^2} )  $,
which completes the computation of the roll amplitude in terms
of the distance to threshold.

\setcounter{equation}{0}
\section{The K\"uppers-Lortz instability}

We assume that the steady rolls of wavevector ${\vec k_1}$ computed
in Section 2 are subject to a perturbation $\tilde{X}$ in 
the form of rolls with an  infinitesimal amplitude and a wavevector
${\vec k_2}$  
making with ${\vec k_1}$  an angle $\theta$ that it is enough to
consider in  the range $] -\frac{\pi}{2} ,\frac{\pi}{2}]$.
We assume for the sake of simplicity that the wavenumbers $|{\vec k_1}|$ 
and $|{\vec k_2}| $
are critical.  
When real, the growth rate $\sigma$  of this perturbation is given by
\begin{equation} (U +R_a  G) \tilde{X} = Q(X,\tilde{X}) +
  Q(\tilde{X},X)+ \sigma  V \tilde{X} . \label{eq: li}
\end{equation} 
In order to compute $\sigma = \epsilon \sigma_1 +\epsilon^2 \sigma_2
+ \cdots $ perturbatively near  threshold, we also expand 
$\tilde{X} = \tilde{X_1} + \epsilon \tilde{X_2} +
\epsilon^2 \tilde{X_3}  + \cdots   \label{eq:ex}$
or, for the individual components,
\begin{eqnarray}
&&\tilde{\phi}  =   \tilde{\phi}_1 \cos\pi z  +
 \epsilon \tilde{\phi}_2 \sin 2\pi z +\cdots  \\
&&\tilde{\psi}  =   \tilde{\psi}_1 \sin\pi z  +
 \epsilon (\tilde{\psi}_0 +\tilde{\psi}_2 \cos 2\pi z )  +\cdots\\
&& \tilde{\vartheta}  =   \tilde{\vartheta}_1 \cos\pi z  +
 \epsilon \tilde{\vartheta}_2 \sin 2\pi z  +\cdots.
\end{eqnarray}
Equation (\ref{eq: li}) leads to
\begin{eqnarray}  
&& L \tilde{X_1}  =  0  \label{eq:li0}  \\
&&  L \tilde{X_2}  =  Q(X_1,\tilde{X_1}) + Q(\tilde{X_1},X_1) +
 \sigma_1  V \tilde{X_1}  \label{eq:li1} \\
&& L \tilde{X_3}  =  Q(X_2,\tilde{X_1}) + Q(\tilde{X_1},X_2) +
Q(X_1,\tilde{X_2}) + Q(\tilde{X_2},X_1)  \nonumber \\
&&+\sigma_1  V \tilde{X_2} +\sigma_2  V \tilde{X_1}  -
 R_2  G \tilde{X_1} \label{eq:li2}    . 
\end{eqnarray}
Using a notation similar to (\ref{eq:notation}), we define 
\begin{equation}
Q(X_i,\tilde{X}_j)+Q(\tilde{X}_j,X_i) = Q_{i,\tilde{j}} =
 (Q^{(1)}_{i ,\tilde{j}}, Q^{(2)}_{i ,\tilde{j}}, Q^{(3)}_{i
   ,\tilde{j}})^t  .  
\end{equation}
Writing the solution of eq. (\ref{eq:li0}) in the form
$ \tilde{X_1} =B v({\vec k_2})+c.c.$, where $B$ is an arbitrary
constant, we have in the right hand side of  eq. (\ref{eq:li1}),
\begin{equation}
Q_{ 1,\tilde{1} } =
\left (\begin{array}{c}
\frac{j_1}{P_r} \delta_h^+
 \delta_h^- (A B^* e^{i {\vec k}^- .{\vec x} } +
 A B e^{i {\vec k}^+ .{\vec x} }) \sin 2 \pi z + c.c. \\
\frac{j_2}{P_r} ( \delta_h^- A B^* e^{i {\vec k}^-.{\vec x} } +
\delta_h^+ A B e^{i {\vec k}^+ .{\vec x}}) + c.c.  \\
j_3  (\delta_h^+ A B^* e^{i {\vec k}^-.{\vec x }} +
\delta_h^- A B e^{i {\vec k}^+ .{\vec x} } ) \sin 2 \pi z + c.c.
  \end{array} \right ) , 
\label{eq:qx1x1}
\end{equation}
where we have introduced the wavevectors
\begin{equation}
{\vec k}^\pm =  {\vec k_1} \pm {\vec k_2}    , 
\end{equation}
and defined the numerical constants
\begin{eqnarray}
 j_1 =   - \frac{3}{2}\pi k^2   , \ \
 j_2  =  -\frac{\tau\pi^2 k^2}{k_p^2}   ,\ \ 
 j_3  =   \frac{R_0 k^2 \pi}{2 k_p^2}  . \label{eq:j}  
\end{eqnarray}
Furthermore, the coefficient $\delta_h^\pm$, given by
 $ \Delta_h e^{ i {\vec k}^\pm . {\vec x}} =
 \delta_h^\pm  e^{i {\vec k}^\pm . {\vec x}}$,
read
\begin{equation}
\delta_h^+ =-4 k^2 \cos^2 \frac{\theta}{2} ,~~~\delta_h^- =
 -4 k^2 \sin^2 \frac{\theta}{2}.  \label{eq:delta}
\end{equation}
Since $\langle v({\vec k_2}), Q(X_1,\tilde{X_1}) +
Q(\tilde{X_1},X_1) \rangle = 
 0$, while $\langle v({\vec k_2}), V \tilde{X_1} \rangle \neq 0$,
 the solvability condition for eq. (\ref{eq:li1}) implies 
 $\sigma_1 =0$. Straightforward algebra then leads to
\begin{eqnarray} 
&& P \Delta_h \tilde{\phi_2}  =  \Delta {Q}^{(1)}_{
    1,\tilde{1} } + 
\tau \partial_z  {Q}^{(2)}_{ 1,\tilde{1} } + \Delta_h {Q}^{(3)}_{
  1,\tilde{1} } \label{eq:livd1}  \\ 
&&  \Delta    \Delta_h \tilde{\psi_2}  =   - 
\tau \partial_z \Delta_h \tilde{\phi_2}  \label{eq:livd2} \\
&&   \Delta    \Delta_h \tilde{\psi_0}  =      {Q}^{(2)}_{
             1,\tilde{1} } \label{eq:livd2bis} \\
&&   \Delta \tilde{\vartheta_2}  =    {Q}^{(3)}_{
             1,\tilde{1} } + 
 R_0 \Delta_h \tilde{\phi_2}.  \label{eq:livd3}
\end{eqnarray}
Solving in the form
\begin{eqnarray}  
&& \tilde{\phi}_2 =  \tilde{\Phi}_2^+ A B
 e^{i {\vec k}^+. {\vec x}}  + \tilde{\Phi}_2^-  A B^* e^{i {\vec k}^-. 
{\vec x}}  + c.c.  \\
&& \tilde{\psi}_2 =  \tilde{\Psi}_2^+ A B
 e^{i {\vec k}^+. {\vec x}}  + \tilde{\Psi}_2^- A B^*  e^{i {\vec
     k}^-. {\vec x}} + c.c. \\ 
&& \tilde{\psi}_0  =  \tilde{\Psi}_0^+  A B e^{i {\vec k}^+. {\vec
    x}} + \tilde{\Psi}_0^- A B^* e^{i {\vec k}^-. {\vec x}} + c.c.  \\
&& \tilde{\vartheta}_2  =     \tilde{\Theta}_2^+ A B
 e^{i {\vec k}^+. {\vec x}}  + 
\tilde{\Theta}_2^-  A B^* e^{i {\vec k}^-. {\vec x}}  
  + c.c.,
\end{eqnarray}
we get\par\noindent
{\begin{tabular}{ll}

$ \tilde{\Phi}_2^{+}   =  \frac{\delta_h^{-}}{p^{+}}
(\frac{j_1}{P_r} \delta^{+} +  j_3 ) $ &
$\tilde{\Phi}_2^{-}   =  \frac{\delta_h^{+}}{p^{-}} (\frac{j_1}{P_r}
\delta^{-} +  j_3 ) $  \\
$ \tilde{\Psi}_2^+     =  -2 \pi \tau \frac{\delta_h^{-}}{p^{+}
  \delta^{+}}  (\frac{j_1}{P_r}  \delta^{+} + j_3 ) $ &
$\tilde{\Psi}_2^-     =  -2 \pi \tau \frac{\delta_h^{+}}{p^{-}
  \delta^{-}}  (\frac{j_1}{P_r}  \delta^{-} + j_3 )$   \\
$ \tilde{\Psi}_0^+     =  \frac{j_2}{P_r \delta_h^+}  
  \label{eq:psi01b}$ &
$\tilde{\Psi}_0^-     =  \frac{j_2}{P_r \delta_h^-}  
 \label{eq:psi02}$ \\
$\tilde{\Theta}_2^+  =  R_0 \frac{\delta_h^{+} \delta_h^{-}}{p^{+}
 \delta^{+}} (\frac{j_1}{P-r}  \delta^{+} + j_3 ) +
 j_3 \frac{\delta_h^{-}}{\delta^{+}} $ &
$\tilde{\Theta}_2^-  =
  R_0 \frac{\delta_h^{+} \delta_h^{-}}{p^{-} \delta^{-}}
  (\frac{j_1}{P_r}  \delta^{-} 
 + j_3 ) +
j_3 \frac{\delta_h^{+}}{\delta^{-}}  $   .  

\end{tabular} \par
\medskip
\noindent The coefficient $\delta^\pm$ and $p^\pm$ defined by the condition
 $ \Delta e^{ i {\vec k}^\pm . {\vec x}} T(2 \pi z) =
 \delta^\pm  e^{i {\vec k}^\pm . {\vec x}} T(2 \pi z)$ 
and $P  e^{ i {\vec k}^\pm . {\vec x}} T(2 \pi z) =
 p^{\pm}  e^{i {\vec k}^\pm . {\vec x}} T(2 \pi z)$, (where the
 function T stands  for sine or cosine), are given by 
\begin{eqnarray}
&& \delta^+ = - [4 \pi^2 + 4 k^2  \cos^2 \frac{\theta}{2}  ]  , \\
&& \delta^- =  - [4 \pi^2 + 4 k^2  \sin^2 \frac{\theta}{2}  ]   ,
\label{eq:dpm} \\ 
&& p^+ = - [4 \pi^2 + 4 k^2  \cos^2 \frac{\theta}{2}  ]^3 - 4 \pi^2
\tau^2 + 4  k^2 R_0   \cos^2 \frac{\theta}{2} ,\label{eq:pp} \\
&&  p^- = - [4 \pi^2 + 4 k^2  \sin^2 \frac{\theta}{2} ]^3 - 4 \pi^2
 \tau^2 + 4 k^2 R_0 \sin^2 \frac{\theta}{2} \label{eq:pm} ,
\end{eqnarray}
where the $\cos^2 \frac{\theta}{2}$ and $\sin^2 \frac{\theta}{2}$
contributions result from the action of the horizontal Laplacian on 
$ e^{i {\vec k}^+ . {\vec x}}$ and $ e^{i {\vec k}^- . {\vec x}}$
respectively. 

An important observation is that the contribution 
$\tilde{\Psi}_0^- A B^*  e^{i {\vec k}^- . {\vec x}}$ to
$\tilde{\psi}_2$ (which disappears at infinite Prandtl number)
diverges in the limit $\theta \rightarrow 0$,  where  it can be
viewed as associated to a ``mean flow'' generated by the rotation.
This term is specific to free-slip  boundary conditions
and has no equivalent  when rigid boundaries are considered. 
The divergence originates from the fact that in
eqs. (\ref{eq:livd1})-(\ref{eq:livd3}),
the dynamics of the mean flow is slaved to that
of the leading convective mode.
This ``adiabatic approximation'' is valid at finite $\theta$ but
breaks down in an ``angular boundary layer'' near $\theta=0$, where 
time derivatives become relevant.
Postponing to Section 4 the analysis of this layer, we
derive here the solvability condition of
eq. (\ref{eq:li2}) for finite $\theta$, by writing
\begin{eqnarray}
&& \langle v({\vec k_2}), G  \tilde{X_1}   \rangle  =
 \frac{1}{2} c_1 c_3 k^2 B^*\\
&& \langle v({\vec k_2}), V \tilde{X_1}   \rangle  =
\frac{1}{2}  (R_0 {P_r}^{-1} c_1^2 k^2 {k_p}^2 - R_0 {P_r}^{-1}
c_2^2 k^2 +  c_3^2)  B^* \\
&& \langle v({\vec k_2}), ( Q(X_2,\tilde{X_1}) + Q(\tilde{X_1},X_2)
\rangle  =\frac{1}{4}  c_1^2 c_3^2 k^4 |A|^2 B^*.
\end{eqnarray}
Furthermore
\begin{equation}
Q(X_1,\tilde{X}_2) + Q(\tilde{X}_2, X_1) =
 Z(q^{(1)}_{ 1,\tilde{2} }, q^{(2)}_{ 1,\tilde{2} }, q^{(3)}_{
   1,\tilde{2} }) |A|^2 B^*  e^{- i {\vec k_2}.{\vec x}} + c.c. + F_3,
\end{equation}
where $F_3$ refers to  non resonant terms proportional to $\sin 3
\pi z$ or $\cos 3 \pi z$. We also have
\begin{eqnarray} 
 q^{(1)}_{ 1,\tilde{2} } & = & \frac{ k^4}{2 P_r}  \sin \theta
 [(-\tilde{\Psi}_2^{+} +2 \tilde{\Psi}_0^{-} ) 
(c_1 \pi^2- c_1 k^2-2 c_1 \pi^2  \cos \theta-2  c_2 \pi  \sin
\theta) \nonumber \\ 
&& + (\tilde{\Psi}_2^{-} -2 \tilde{\Psi}_0^{+} )
 (c_1 \pi^2-c_1 k^2+2 c_1 \pi^2  \cos \theta+2 c_2 \pi  \sin \theta)
 \nonumber \\ 
&& + 2 \tilde{\Phi}_2^{-}  k_p^2 (c_2 (1- \cos \theta)+\pi c_1
\sin \theta)  
- 2 \tilde{\Phi}_2^{+}  k_p^2 (c_2 (1+ \cos \theta)-\pi c_1  \sin
\theta) ]  \nonumber \\ 
&&\\
q^{(2)}_{ 1,\tilde{2} } & = & -\frac{ k^4}{2 P_r}[2
\tilde{\Psi}_0^{-}   
(c_1 \pi (-1+2  \cos \theta-\cos 2 \theta)+c_2 ( \sin \theta-\sin 2
\theta)) \nonumber \\ 
&& + 2 \tilde{\Psi}_0^{+}  
(c_1 \pi (-1-2  \cos \theta-\cos 2 \theta)-c_2 ( \sin \theta+\sin 2
\theta)) \nonumber \\ 
&& +\tilde{\Psi}_2^{+}   \sin \theta (-2 c_1 \pi  \sin \theta+c_2
(1+2  \cos \theta))\nonumber \\ 
&& +\tilde{\Psi}_2^{-}   \sin \theta (-2 c_1 \pi  \sin \theta-c_2
  (1-2  \cos \theta))\nonumber \\ 
&& +\tilde{\Phi}_2^{+}  (c_2 \pi (1-\cos 2 \theta)-\pi^2 c_1 (2
\sin \theta-\sin 2 \theta))\nonumber \\ 
&& +\tilde{\Phi}_2^{-}  (c_2 \pi (1-\cos 2 \theta)+\pi^2 c_1 (2
  \sin \theta+\sin 2 \theta))  ]  \\ 
 q^{(3)} _{ 1,\tilde{2} } & = &   \frac{k^2}{2} [ \tilde{\Theta_2}^- 
(c_1 \pi (1+ \cos \theta)+c_2  \sin \theta) 
+ \tilde{\Theta_2}^+ (c_1 \pi (1- \cos \theta)-c_2  \sin \theta)]
\nonumber \\ 
&& +  \frac{k^2}{2} c_3 [\tilde{\Psi}_2^{+} +2 \tilde{\Psi}_0^{+} 
-\tilde{\Psi}_2^{-} -2 \tilde{\Psi}_0^{-} ]  \sin \theta  .
\end{eqnarray}
It follows that
\begin{equation}
\langle v({\vec k_2}), Q(X_1,\tilde{X_2}) + Q(\tilde{X_2},X_1)
\rangle =  
\frac{1}{2} (R_0 c_1 q^{(1)}_{1,\tilde{2} } + 
 R_0 c_2  q^{(2)}_{ 1,\tilde{2} }+c_3 q^{(3)}_{ 1,\tilde{2} } )
 |A|^2 B^*,  \label{eq:q123} 
\end{equation}
and finally
\begin{equation}
\sigma = \epsilon^2 \sigma_2 = \epsilon^2  \frac{r_2  c_1 c_3 k^2 -
  (\frac{1}{2}  c_1^2 c_3^2 k^4 + 
 R_0 c_1 q^{(1)}_{ 1,\tilde{2} } +  R_0 c_2 q^{(2)}_{ 1,\tilde{2} } 
+ c_3 q^{(3)}_{ 1,\tilde{2} } )}
{ R_0 {P_r}^{-1} c_1^2 k^2 {k_p}^2 - R_0 {P_r}^{-1} c_2^2 k^2 +
 c_3^2}  |A|^2  ,    \label{eq:sigout}
\end{equation}
where $\epsilon^2 |A|^2$ can  be expressed 
as $\frac{2 k_p^2}{R_0 k^4 -\frac{\tau^2 \pi^4}{P_r^2}} (R_a - R_0)$.

Since in the limit $\theta \rightarrow 0$, 
 $\tilde{\Psi}_0^{-}$ diverges like $\sin^{-2} \frac{\theta}{2}$,
the quantity  $q^{(l)}_{1,\tilde{2}} $ with $l=1, 2, 3$,
scales like $ \sin\frac{\theta}{2} \tilde{\Psi}_0^{-} \sim \sin^{-1}
\frac{\theta}{2}$, 
and the growth rate behaves like
\begin{equation}
\sigma \sim \frac{\tau \pi^2 k^2 |A|^2 }{2 k_p^2 P_r}  \frac{
  \epsilon^2}{\sin\frac{\theta}{2}} ,
\label{eq:diverg0}
\end{equation} 
indicating a breakdown of the above asymptotics at finite Prandtl
numbers, in the case of small angle perturbations.

Pushing the $\theta$-expansion at the next order, (as needed in
Section 4), we write
\begin{equation}
\sigma \sim [- \eta +\frac{\tau \pi^2 k^2}{2 k_p^2 P_r} (2 \xi
+\frac{ 1}{\sin\frac{\theta}{2}} )] \epsilon^2 |A|^2,   
\label{eq:diverg}
\end{equation} 
with 
\begin{equation}
\eta =  \frac{r_2 }{R_0}  \frac{k_p^2}{(1 + \frac{1}{P_r}
(1 -\frac{2 \tau^2 \pi^2}{R_0 k^2}) )}  
 \label{eq:eta}
\end{equation}
 and
\begin{equation}
\xi =- \frac{ \tau \pi^2}{P_r k_p^2 k^2 (1 + \frac{1}{P_r} (1
  -\frac{2 \tau^2 
\pi^2}{R_0 k^2}))} ,
\label{eq:xi}
\end{equation}
the latter coefficient collecting contributions originating from 
$\tilde{\Psi}_0^{-}$.

The  divergence shown in eq. (\ref{eq:diverg0})  was noted in
\cite{Knobloch}.  
It indicates that the above analysis should be viewed as an outer
expansion, and that a different scaling is required for small $\theta$.
In the following, the growth rate given by eq. (\ref{eq:sigout}) will
thus be denoted $\sigma_{outer}$.

\setcounter{equation}{0}
\section{The small-angle instability}
The small angle divergence of the stream function 
$\psi_0 \sim \epsilon  \sin^{-2}\frac{\theta}{2} $ 
 and of the growth rate $\sigma_{outer}$
$\sim \epsilon^2 \sin^{-1}\frac{\theta}{2} $, 
indicates that new scalings in $\epsilon$ are expected in an
angular boundary layer near $\theta=0$.
Denoting by $\epsilon^\alpha $ the thickness of this layer, by
 $\epsilon^\beta$ the amplitude of $\Psi_0$ and by
 $\epsilon^\gamma$ the magnitude of the growth rate in this layer,
 the matching of the ``outer'' and ``inner'' regions requires
$\beta = 1- 2\alpha$ and $\gamma = 2 - \alpha$. Since,
in the inner region, the time derivative in the mean flow equation
(whose presence will remove the divergence) becomes
comparable to the viscous term when  $\gamma = 2 \alpha$, 
we get
$\alpha = \frac{2}{3}$, $\beta = -\frac{1}{3}$ and $\gamma =
\frac{4}{3}$. 

Furthermore, when expanding  eq. (\ref{eq:mb1}) inside 
the boundary layer, the parameter $\epsilon$
appears not only through the horizontal 
Fourier modes of $X_1$ whose amplitudes scale like 
entire powers of $\epsilon$, but also through 
the angular dependence of the operators involved
in this equation. We are thus led to expand 

\begin{eqnarray}
&& \sigma = \epsilon \sigma_1 +  \epsilon^{\frac{4}{3}}
\sigma_{\frac{4}{3}} + \epsilon^{\frac{5}{3}}  \sigma_{\frac{5}{3}}
+ \epsilon^2 \sigma_2 +  
\epsilon^{\frac{7}{3}}  \sigma_{\frac{7}{3}} +\epsilon^{\frac{8}{3}}
\sigma_{\frac{8}{3}}+\cdots \label{eq:sigmatilde} 
\end{eqnarray}
and
\begin{eqnarray}
&& \tilde{X} = \epsilon^{-\frac{1}{3}} \tilde{Y}_{-\frac{1}{3}} 
+\tilde{Y}_0 + \tilde{X}_{1} 
+ \epsilon^{\frac{1}{3}} \tilde{Y}_{\frac{1}{3}}
+ \epsilon^{\frac{2}{3}} \tilde{Y}_{\frac{2}{3}}
+\epsilon \tilde{X}_{2} +
 \epsilon^{\frac{4}{3}} \tilde{X}_{\frac{4}{3}} 
+ \epsilon^{\frac{5}{3}} \tilde{X}_{\frac{5}{3}} \nonumber  \\ 
&& + \epsilon^2 \tilde{X}_{3}+ 
\epsilon^{\frac{7}{3}} \tilde{X}_{\frac{7}{3}}
+\epsilon^{\frac{8}{3}} \tilde{X}_{\frac{8}{3}} 
 + \epsilon^3 \tilde{X}_{4}+\cdots   \label{eq:xtilde} 
\end{eqnarray}
where terms of the form $\tilde{Y}_{\mu} = \tilde{\Psi}_{\mu}
B^* A  e^{i {\vec k}^- . {\vec x}}   
(0 , 1, 0)^t + c.c. $, are introduced to
cancel almost resonant contributions resulting from the interaction of
the basic rolls with quasi-parallel perturbations.
As seen later, in the boundary layer, $\sigma$ can be complex.

Substituting (\ref{eq:xtilde}) and (\ref{eq:sigmatilde}) in
eq. (\ref{eq: li}) and concentrating on perturbations such that the
angle $\theta$ between the wavevectors $\vec{k}_1$ and $\vec{k}_2$
is of order $\epsilon^{\frac{2}{3}}$, we obtain the following
hierarchy.
\par
$\bullet$ At order $\epsilon^0$,
\begin{equation}
 L \tilde{X}_{1} =0 \label{eq:tdv0} ,
\end{equation}
leading to
\begin{equation}
\tilde{X}_1 = v({\vec k}_2) B + c.c.  .
\end{equation}
\par
$\bullet$ At order $\epsilon$,
\begin{eqnarray}
L \tilde{X_2}=   \left ( \begin{array}{c} 
  0 \\ 
- \frac{4 j_2 k^2}{P_r}  A B e^{i {\vec k}^+. {\vec x}} \\ 
- 4 j_3 k^2  A B^* e^{i {\vec k}^-. {\vec x }} \sin 2 \pi z  
\end{array} \right) + 
\sigma_1 B V v({\vec k}_2) + c.c.\ .
\label{eq:tdv1}
\end{eqnarray}
The solvability condition reads 
\begin{equation}
\epsilon \sigma_1 = 0,  \label{eq:s1}
\end{equation} 
and the solution is given by
\begin{equation}
\tilde{X_2} = \left ( \begin{array}{c}
0   \\ 
 \tilde{\Psi}_0^+ A B e^{i {\vec k}^+. {\vec x}}  \\
  \tilde{\Theta_2}^-  A B^* e^{i {\vec k}^-. {\vec x}} \sin 2 \pi z 
\end{array} \right ) + c.c.  
\end{equation} 
with $\tilde{\Psi}_0^+=  -\frac{j_2}{4 k^2 P_r}$ and 
$\tilde{\Theta}_2^-  =   j_3 \frac{k^2}{\pi^2}$.
\par
$\bullet$ At order $\epsilon^{\frac{4}{3}}$,
\begin{eqnarray} 
&&\epsilon^{\frac{4}{3}} L \tilde{X}_{\frac{4}{3}}  =
\epsilon^{\frac{2}{3}} \sin\theta 
\tilde{\Psi}_{-\frac{1}{3}}^* |A|^2 B e^{ i\vec{k_2} .\vec{x}} 
Z(- \frac{k^4 k_p^2}{P_r} \ , \frac{k^4 c_2}{P_r}  \ ,
- k^2 c_3)   \nonumber\\
& &\quad\quad + \epsilon^{\frac{4}{3}} \sigma_{\frac{4}{3}} V  B
v({\vec k_2}) + c.c.+ \cal{NR}, \label{eq:f43}
\end{eqnarray} 
where $\cal{NR}$ collects non-resonant terms. 
The solvability condition is
\begin{equation}
\epsilon^{\frac{4}{3}} \sigma _{\frac{4}{3}} = 
\epsilon  k^2 \sin\theta |A|^2 
\epsilon^{-\frac{1}{3}} \tilde{\Psi}_{-\frac{1}{3}}^{*}.
\label{eq:sig}
\end{equation}
\par
$\bullet$ At order $\epsilon^{\frac{5}{3}}$,
\begin{eqnarray}
&&\epsilon^{\frac{5}{3}}  L \tilde{X}_{\frac{5}{3}} = 
\epsilon \sin\theta \tilde{\Psi}_{0}^* |A|^2 B   e^{ i\vec{k_2}
  .\vec{x}}  
Z(- \frac{k^4 k_p^2}{P_r}  \ , \frac{k^4 c_2}{P_r}   \ ,
- k^2 c_3 )\nonumber\\
&& \quad\quad + \epsilon^{\frac{5}{3}} \sigma_{\frac{5}{3}} V  B
v({\vec k_2}) + c.c.+ \cal{NR} . \label{eq:f53} 
\end{eqnarray}
The solvability of this equation requires
\begin{equation}
 \epsilon^{\frac{5}{3}} \sigma _{\frac{5}{3}} = 
\epsilon k^2 \sin{\theta} |A|^2
 \tilde{\Psi}_{0}^{*}.    \label{eq:sig53}
 \end{equation}
\par
$\bullet$ At order $\epsilon^{2}$,
\begin{eqnarray}
 && \epsilon^2 L \tilde{X_3} =  \epsilon^2 Q_{\tilde{1},2} +
 \epsilon^2  Z( \bar{q}^{(1)}_{1,\tilde{2} },
 \bar{q}^{(2)}_{1,\tilde{2} },  
\bar{q}^{(3)}_{1,\tilde{2} })  |A|^2  B e^{ i\vec{k_2} .\vec{x}}
\nonumber \\ 
& & \quad\quad + \epsilon^{\frac{2}{3}} \sin\theta
\tilde{\Psi}_{-\frac{1}{3}}^{*}  e^{ i\vec{k_2} .\vec{x}}|A|^2  B
Z( -2 \frac{k^4}{P_r} c_2 \pi \sin\theta  \ ,
 \frac{k^4}{P_r} c_1 \pi(\sin \frac{\theta}{2} - \sin \frac{3
  \theta}{2}) \ ,0 )  
 \nonumber \\
& &\quad\quad + \epsilon^{\frac{4}{3}} \sin\theta
\tilde{\Psi}_{\frac{1}{3}}^* |A|^2 B e^{ i\vec{k_2} .\vec{x}}  
Z(- \frac{k^4 k_p^2}{P_r}, \frac{k^4 c_2}{P_r},- k^2 c_3)
  \nonumber \\ 
& & \quad\quad - \epsilon^2 R_2 G \tilde{X_1} 
    +   \epsilon^2 \sigma_2  V  B v({ \vec k_2})+  c.c. + \cal{NR},
   \label{eq:tdv2}  
\end{eqnarray}
where
\begin{equation}
  ({ \bar{q}}^{(1) }_{1,\tilde{2} }, 
{ \bar{q}}^{(2) }_{1,\tilde{2} },{ \bar{q}}^{(3) }_{1,\tilde{2} }) =
 (0, 4 \frac{k^4}{ P_r} c_1 \pi  \tilde{\Psi}_0^+ , k^2 c_1 \pi
 \tilde{\Theta}_2^-) \label{eq:q123bis} 
\end{equation}
denotes the limit as $\theta \rightarrow 0$ of the vector $( {q}^{(1)
  }_{1,\tilde{2} },  
 {q}^{(2) }_{1,\tilde{2} }, {q}^{(3) }_{1,\tilde{2} })$
 from which the contributions coming from
 $\Psi_0^-$ have been removed. The solvability condition reads 
\begin{eqnarray}
 \epsilon^2 \sigma_2 =-\eta|A|^2 \epsilon^2
+ 2\epsilon^{\frac{2}{3}}  \xi  k^2 \sin\theta
\sin\frac{\theta}{2}  |A|^2  
\tilde{\Psi}_{-\frac{1}{3}}^{-*} 
+ \epsilon^{\frac{4}{3}} k^2 \sin\theta |A |^2
 \tilde{\Psi}_{\frac{1}{3}}^{-*}  ,
\label{eq:scsigma2}
\end{eqnarray}
where
\begin{equation}
\eta = - \frac{r_2  c_1 c_3 k^2 - (\frac{1}{2}  c_1^2 c_3^2 k^4 +
 R_0 c_1  {\bar{q}}^{(1) }_{1,\tilde{2} }
+  R_0 c_2 {\bar{q}}^{(2) }_{1,\tilde{2} } +c_3 {\bar{q}}^{(3)
  }_{1,\tilde{2} } ) } 
{ R_0 {P_r}^{-1} c_1^2 k^2 {k_p}^2 - R_0 {P_r}^{-1} c_2^2 k^2 +
 c_3^2}    
\end{equation}
identifies with the expression given in eq. ( \ref{eq:eta}). The
coefficient  $\xi$ is given by eq. (\ref{eq:xi}).

Combining the solvability conditions 
(\ref{eq:s1}), (\ref{eq:sig}), (\ref{eq:sig53}) and
(\ref{eq:scsigma2}), we are led to express the growth rate
\begin{eqnarray}
\sigma_{inner} &=&\epsilon^{\frac{4}{3}} \sigma_{\frac{4}{3}} +
\epsilon^{\frac{5}{3}} \sigma_{\frac{5}{3}} + \epsilon^{2} \sigma_{2},
\end{eqnarray}
in terms of the ``mean flow'' 
\begin{eqnarray}
\Psi&=& \epsilon^{-\frac{1}{3}} \tilde{\Psi}_{-\frac{1}{3}}^{*} 
+ \tilde{\Psi}_{0}^{*} +
 \epsilon^{\frac{1}{3}} \tilde{\Psi}_{\frac{1}{3}}^{*},
\end{eqnarray}
in the form 
\begin{eqnarray}
\sigma_{inner} = \epsilon k^2 \sin\theta |A|^2 (1 + 2 \xi
\sin\frac{\theta}{2}) \Psi - \epsilon^2 \eta |A|^2,
\label{eq:eqsiginner} 
\end{eqnarray}
where subdominant corrections have been neglected.

In order to estimate the mean flow $\Psi$, we  push the
$\epsilon$-expansion of  eq. (\ref{eq:mb1}) at higher orders, where
the beating of  
the perturbation with the basic solution produces contributions of
the form $e^{i ({\vec k}_1 - {\vec k}_2).{\vec x}}$ which become
space-independent and thus resonant in the  small
$\theta$ limit. Consequently, uniform boundedness of the solutions 
requires, in addition to the usual solvability conditions, the prescription
of ``quasi-solvability conditions'' aimed to eliminate terms which are
strictly resonant only for $\theta = 0$. This approach is
similar to that used by Ablowitz and Benney \cite{Abl}
when dealing with the small-amplitude divergence of the Whitham
modulation analysis for nonlinear dispersive waves (see also
\cite{Newell}). 
These authors modify the (algebraic) dispersion relation by means of
additional corrective terms determined by a constraint which becomes
an actual solvability condition in the small amplitude limit, thus
transforming the algebraic dispersion relation arising in Whitham's theory,
into a partial differential 
equation for the wave amplitude. In the context of rotating
convection, we include contributions $\tilde Y_n$ proportional to
$ e^{i({\vec k_1}-{\vec k_2}).{\vec x}}$ in the perturbation 
expansion, which are determined by cancelling them with the
terms displaying the same functional dependence and 
originating from the beating of the basic rolls with
quasi-parallel perturbations. Like in the small-amplitude limit of
nonlinear waves, 
this condition becomes an actual solvability in the limit $\theta
\rightarrow 0$. In both instances, the singularity is prevented by
removing slaving conditions: that of the amplitude
with respect to the phase in the case of waves, or that of the mean
flow with respect to the convective modes in the present problem
(compare eqs. (\ref{eq:livd2}) and (\ref{eq: free}) below).
\par
$\bullet$ At  order $\epsilon^{\frac{7}{3}}$,
\begin{eqnarray}
&&\epsilon^{\frac{7}{3}} L \tilde{X}_{\frac{7}{3}} +
 \left ( \begin{array}{c} 
\epsilon \delta_h^-  \tilde{\Theta}_2^-  A B^* e^{i {\vec k}^-.{\vec
    x} } \sin 2 \pi z\\ 
\epsilon^{-\frac{1}{3}} {\delta_h^- }^2  \tilde{\Psi}_{-\frac{1}{3}}
 A B^* e^{i {\vec k}^-. {\vec x }} + c.c. \\  
0 
\end{array} \right )  =   \nonumber \\
&& \epsilon^{\frac{7}{3}} [Q_{1,\tilde{\frac{4}{3}}} ] + 
    \epsilon  
\left (\begin{array}{c}
\frac{j_1}{P_r} \delta_h^+ \delta_h^- 
(A B^* e^{i {\vec k}^- .{\vec x} } + A B e^{i {\vec k}^+ .{\vec x} }
) \sin 2 \pi z + c.c. \\ 
\frac{j_2}{P_r}  \delta_h^- A B^* e^{i {\vec k}^-.{\vec x} } + c.c.  \\
j_3   \delta_h^- A B e^{i {\vec k}^+ .{\vec x} } \sin 2 \pi z + c.c.  
\end{array} \right )  + \nonumber \\
 &&    \left ( \begin{array}{c}  0  \\
\epsilon^{\frac{4}{3}} \sigma^*_{\frac{4}{3}} 
P_r^{-1}  \delta^-_h  \epsilon^{-\frac{1}{3}} 
\tilde{\Psi}_{-\frac{1}{3}}^{}  A B^* e^{i {\vec k}^-. {\vec x }} +
c.c. \\ 
  0 \end{array} \right )   +\nonumber \\
&& \epsilon \sin\theta
\tilde{\Psi}_{0}^{*}  e^{ i\vec{k_2} .\vec{x}}|A|^2  B
Z( -2 \frac{k^4}{P_r} c_2 \pi \sin\theta  \ ,
 \frac{k^4}{P_r} c_1 \pi(\sin \frac{\theta}{2} - \sin \frac{3
  \theta}{2}) \ ,0 ) + \nonumber \\ 
&& + \epsilon^{\frac{5}{3}} \sin\theta
\tilde{\Psi}_{\frac{2}{3}}^* A   e^{ i\vec{k_2} .\vec{x}}  
Z(- \frac{k^4 k_p^2}{P_r}, \frac{k^4 c_2}{P_r},- k^2 c_3) +
 \epsilon^{\frac{7}{3}}  \sigma_{\frac{4}{3}} V \tilde{X}_2 +
\epsilon^{\frac{7}{3}} \sigma_{\frac{7}{3}} V \tilde{X}_1 ,\nonumber \\
\label{eq:tdv73}  
\end{eqnarray}
where $[Q_{1,\tilde{\frac{4}{3}}} ]$ denotes the leading order of 
$Q(X_1,\tilde{X}_{\frac{4}{3}}) + Q(\tilde{X}_{\frac{4}{3}},X_1)$.
Although $\tilde{X}_{\frac{4}{3}}$ contains terms proportional to
$e^{\pm i {\vec k}_2.{\vec x}}$, the resulting contributions of the
form $e^{ i {\vec k}^{-}.{\vec x}}$ in $Q_{1,\tilde{\frac{4}{3}}}$
are preceded by a factor proportional to
$\sin^2\frac{\theta}{2}$ and thus not included in
$[Q_{1,\tilde{\frac{4}{3}}} ]$. 
The quasi-solvability condition thus reads
\begin{equation}
 -\epsilon \sigma_{\frac{4}{3}} P_r^{-1} \delta_h^-
 \tilde{\Psi}_{-\frac{1}{3}}^{*}  +
\epsilon^{-\frac{1}{3}} {\delta_h^-}^2
\tilde{\Psi}_{-\frac{1}{3}}^{*}  - 
\epsilon \delta_h^-  j_2 P_r^{-1} = 0.  \label{eq:solv73}
\end{equation}
\par
$\bullet$ At order $\epsilon^{\frac{8}{3}}$,
\begin{eqnarray}
&&\epsilon^{\frac{8}{3}} L \tilde{X}_{\frac{8}{3}}  
+ \left( \begin{array}{c} 0 \\ -P_r^{-1} \delta_h^-
\epsilon^{\frac{4}{3}} \sigma_{\frac{4}{3}} \tilde{\Psi}^*_0 +
{\delta_h^-}^2 \tilde{\Psi}^*_0 - P_r^{-1} \epsilon^{\frac{4}{3}}
\sigma_{\frac{5}{3}} \tilde{\Psi}^*_{-\frac{1}{3}} \\ 0 \end{array}
\right ) = [Q_{\tilde{\frac{5}{3}},1}] + \nonumber \\
&&\left (\begin{array}{c}
0\\
\epsilon^2  \frac{8}{P_r} k^4 \Psi_0   ( B \epsilon^{-\frac{1}{3}}
 \tilde{\Psi}_{-\frac{1}{3}}^{*}  e^{i {\vec k}^+ .{\vec x} } + c.c. ) 
\sin{\theta} \cos^2\frac{\theta}{2} \\
0 \end{array} \right ) +  \nonumber \\
&& \epsilon^{\frac{4}{3}} \sin\theta
\tilde{\Psi}_{\frac{1}{3}}^{*}  e^{ i\vec{k_2} .\vec{x}}|A|^2  B
Z( -2 \frac{k^4}{P_r} c_2 \pi \sin\theta  \ ,
 \frac{k^4}{P_r} c_1 \pi(\sin \frac{\theta}{2} - \sin \frac{3
  \theta}{2}) \ ,0 ) .\nonumber \\ \label{eq:eq83}
\end{eqnarray}
The quasi-solvability condition is
\begin{eqnarray}
 -P_r^{-1} \epsilon^{-\frac{4}{3}}  \sigma_{\frac{4}{3}}
 \delta_h^- \tilde{\Psi}^*_0  +
{\delta_h^-}^2 \tilde{\Psi}^*_0 - P_r^{-1}
\epsilon^{\frac{5}{3}} \sigma_{\frac{5}{3}} \delta_h^- 
(\epsilon^{-\frac{1}{3}}\tilde{\Psi}^*_{-\frac{1}{3}}) = 0
\label{eq:cs83} 
\end{eqnarray}
where, as previously, $[Q_{\tilde{\frac{5}{3}},1}]$ does not
contribute.
\par
$\bullet$ At order $\epsilon^{3}$,
\begin{eqnarray}
&&\epsilon^3 L \tilde{X}_4  = 
\epsilon^3 [Q_{1,\tilde{3}} + Q_{3,\tilde{1}} +  Q_{2,\tilde{2}}] 
 -\epsilon^3 R_3 G \tilde{X}_1 -\epsilon^3 R_2 G \tilde{X}_2
+ \epsilon^3 ( \sigma_2 V \tilde{X}_2 + \sigma_3 V \tilde{X}_1
)\nonumber \\
&& - (0 \ ,\epsilon^{\frac{1}{3}} {\delta_h^- }^2
\tilde{\Psi_{\frac{1}{3}}^{*}} A B^* e^{-i {\vec k}^-. {\vec x }} +
c.c. \ ,   0  )^t   \nonumber \\ 
&& +  (   0  \ ,P_r^{-1}  \delta^-_h  \epsilon^{\frac{5}{3}}
(\sigma_{2}   \tilde{\Psi}_{-\frac{1}{3}}^{*} +
  \sigma_{\frac{4}{3}} 
 \tilde{\Psi}_{\frac{1}{3}}^{*} + \sigma_{\frac{5}{3}}
 \tilde{\Psi}_{0}^{*} )  A B^* e^{-i {\vec k}^-. {\vec x }} + c.c. \ ,
  0  )^t ,   \label{eq:equate3}
\end{eqnarray}
with the quasi-solvability condition  
\begin{equation}
{\delta_h^- }^2 
(\epsilon^{\frac{1}{3}} \tilde{\Psi}_{\frac{1}{3}}^{-*}) -
P_r^{-1} \epsilon^{2} \sigma_{2}  
 \delta^-_h (\epsilon^{-\frac{1}{3}} \tilde{\Psi}_{-\frac{1}{3}}^{*}) -
P_r^{-1}  \epsilon^{\frac{4}{3}} \sigma_{\frac{4}{3}}  
 \delta^-_h ( \epsilon^{\frac{1}{3}}
 \tilde{\Psi}_{\frac{1}{3}}^{*}) -
 P_r^{-1}  \epsilon^{\frac{5}{3}}  \sigma_{\frac{5}{3}} 
 \delta^-_h \tilde{\Psi}_{0}^{*} = 0. \label{eq:cs93}
\end{equation}

Combining eqs. (\ref{eq:solv73}), (\ref{eq:cs83}) and
(\ref{eq:cs93}), we get, up to subdominant contributions,
\begin{eqnarray}
- P_r^{-1} \sigma_{inner} \Psi + \delta_h^- \Psi = \epsilon P_r^{-1}
j_2 \label{eq: free}
\end{eqnarray}
which together with eq. (\ref{eq:eqsiginner}), constitute a closed
system. 
Solving the resulting quadratic equation for the growth
rate, we obtain two solutions  
\begin{eqnarray}
 &&\sigma^{\pm}_{inner} = \frac{1}{2} (-\epsilon^2 \eta |A|^2 - 4
 k^2 P_r 
   \sin^2\frac{\theta}{2})  \pm  \nonumber \\
&& \frac{1}{2} \left[ (\epsilon^2 \eta |A|^2 - 4 k^2 P_r
   \sin^2\frac{\theta}{2})^2 - 4 \epsilon^2 k^2 \sin\theta |A|^2 j_2 
(1 + 2 \sin\frac{\theta}{2} \xi) \right]^{\frac{1}{2}}  .\nonumber\\ \label{eq:siginner} 
\end{eqnarray}
where $\eta$,  $\xi$ and $j_2$ are defined by eqs. (\ref{eq:eta}),
(\ref{eq:xi}) and (\ref{eq:j}).  This expression covers several regimes

(i) For $\theta \sim \epsilon^{\frac{2}{3}}$,
\begin{equation}
 \sigma^{\pm}_{inner} \sim \frac{1}{2} ( - 4 k^2 P_r
   \sin^2\frac{\theta}{2})  \pm  
 \frac{1}{2} \left[ ( 4 k^2 P_r
   \sin^2\frac{\theta}{2})^2 - 4 \epsilon^2 k^2 \sin\theta |A|^2 j_2 
     \right]^{\frac{1}{2}} 
 \label{eq:siginner-e34} .
\end{equation}
In this range, $ \sigma^{+}_{inner}>0$ for $\theta>0$ ($\theta<0$)
if $\tau >0$ (resp. $\tau <0$) for any finite value of the Prandtl
number (still assuming $P_r > 0.6766$) and of the Taylor number.

(ii) When $\theta \gg \epsilon^{\frac{2}{3}}$,
 \begin{equation}
{\sigma_{inner}^+} \sim \sigma_{match} =
[- \eta+ \frac{\tau \pi^2 k^2}{2 k_p^2 P_r }  
(2 \xi + \frac{1}{\sin\frac{\theta}{2}}) ]
\epsilon^2 |A|^2, \label{eq:siginter}
\end{equation}
and matches the limit of $\sigma_{outer}$ as $\theta \rightarrow 0$.
Similarly, 
\begin{equation}
{\sigma_{inner}^- } \sim - 4 k^2 P_r \sin^2 \frac{\theta}{2}
\end{equation}
is negative and becomes of order unity outside the boundary layer.

(iii) For $\theta \sim \epsilon^2$, 
\begin{eqnarray}
 \sigma^{\pm}_{inner} \sim \frac{1}{2} (-\epsilon^2 \eta |A|^2)  \pm 
 \frac{1}{2} \left[ (\epsilon^2 \eta |A|^2)^2 - 4 \epsilon^2 k^2
   \sin\theta |A|^2 j_2  \right]^{\frac{1}{2}} .  \label{eq:siginner-e2} 
\end{eqnarray}
and for $\theta=0$, $\sigma^{+}_{inner}$ vanishes, while 
$\sigma^{-}_{inner}(0)=-\epsilon^2\eta |A|^2$.

We thus obtain a uniform representation  for  $\theta \in ]
-\frac{\pi}{2},\frac{\pi}{2}]$, of the instability growth
rate near the
convection threshold, of the form
\begin{eqnarray}
\sigma^{+} &=& \sigma_{inner}^{+} + \sigma_{outer} - 
 \sigma_{match}. \label{eq:sunif}
\end{eqnarray}
where the various terms arising in the right-hand-side of
eq. (\ref{eq:sunif}) are given by 
eq. (\ref{eq:sigout}), (\ref{eq:siginner}) and (\ref{eq:siginter}).
The influence of various parameters like the Prandtl number 
and the rotation rate on the stength of the instability, 
is illustrated in the following figures.

Figure 1 shows the variation of the eigenvalues 
$\sigma^{\pm}$ with the angle $\theta$
of the perturbation for $P_r = 2 $, $\epsilon=0.1$ and $\tau=10$.
For anti-clockwise rotation and finite Prandtl number, the growth 
rate $\sigma^{+}$ is positive for small enough positive angles
$\theta$. 
There is also a range of negative angles, where there are two complex
conjugate eigenvalues, with negative real parts. The dashed line
represents the outer solution $\sigma_{outer}$  which diverges in
the limit $\theta \rightarrow 0$. 
The other eigenvalue $\sigma^-$ which, as $\epsilon\rightarrow 0$,
becomes marginal in a neighborhood of $\theta =0$, is of order
unity outside the angular boundary layer. It thus cannot be computed
perturbatively for $\theta$ order unity
but, being always negative or complex with a negative real part,
it cannot lead to an instability.

Figure 2 displays the  growth rate $\sigma^{+}$ for
$\tau=38$, $\epsilon=0.1$ and various values of the  Prandtl
number for positive angles. 
We observe that both the range of unstable angles and the maximal growth
rate decrease when the Prandtl number is increased.
At $P_r =10$, the small angle instability and the K\"uppers-Lortz
instability (around $\theta =50^{o}$) can be separated, in contrast with
the case of smaller Prandtl numbers (e.g. $P_r =5$) where all the angles
 $0 <\theta \leq  64^{o}$ are unstable. For this rotation rate,
only the small angle instability survives at Prandtl number $P_r
=15$. It becomes hardly visible at $P_r=50$.
Indeed, as the Prandtl number goes to infinity, the negative eigenvalue
$\sigma^-$ has a limit, while the outer expansion $\sigma^+_{outer}$
extends towards $\theta =0$ where it asymptotically reaches the
value $\sigma^- (0)$, the inner range reducing to the vertical axis.
         
Figure 3 shows the variation of the instability growth rate with the rotation
rate $\tau$, for $P_r = 15 $ and $\epsilon = 0.1$.
For $\tau =10$, only the small angle instability is present. The
K\"uppers-Lortz instability (again localized around $\theta
=58^{o}$) arises for $\tau \approx 40$ and is strongly amplified  as $\tau$
is increased.  

Figure 4a displays for $\epsilon = 0.1$, the critical value of the
rotation rate $\tau$ for the onset of the K\"uppers-Lortz
instability, as a function of the Prandtl number,
as long as the latter is large enough for the two instabilities to
be separated.
Figure 4b shows the most unstable angle (in degrees) for
the K\"uppers-Lortz instability, versus the Prandtl
number, for a rotation rate corresponding to the onset of the
instability.

\clearpage

\begin{figure}[b]
\centerline{\includegraphics[width=8.6cm]{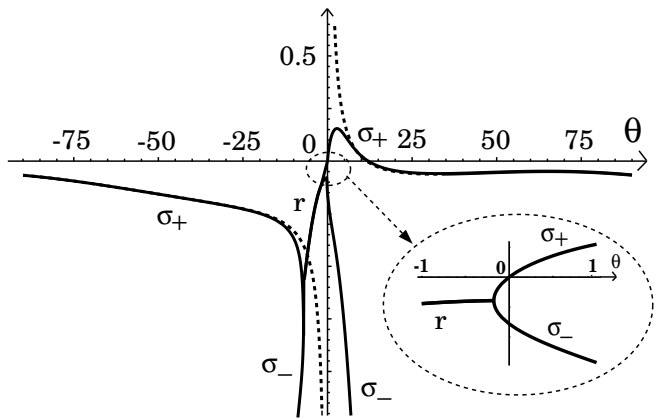}}
\caption{ Instability growth rates $\sigma^+$ and $\sigma^-$
or their real part $r$ when complex conjugate 
 (full line), together with the diverging 
``outer solution'' (dashed line), versus the perturbation angle $\theta$
(in degrees), for $P =10$, $\tau = 10$ and $\epsilon  =0.1$.}
\end{figure}

\begin{figure}[htb]
\centerline{\includegraphics[width=8.6cm]{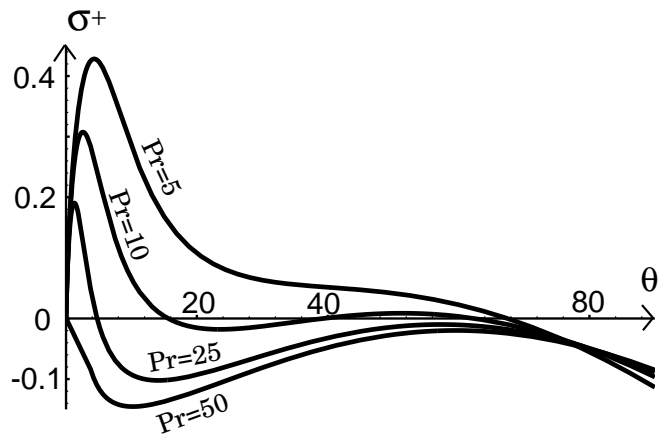}}
\caption{ Growth rate $\sigma^{+}$, 
versus the perturbation angle $\theta >0 $, for $\tau=38$,
$\epsilon=0.1$ and different values $P_r= 5, 10, 25, 50$ of the
Prandtl number.}
\end{figure}

\clearpage

\begin{figure}[htb]
\centerline{\includegraphics[width=8.6cm]{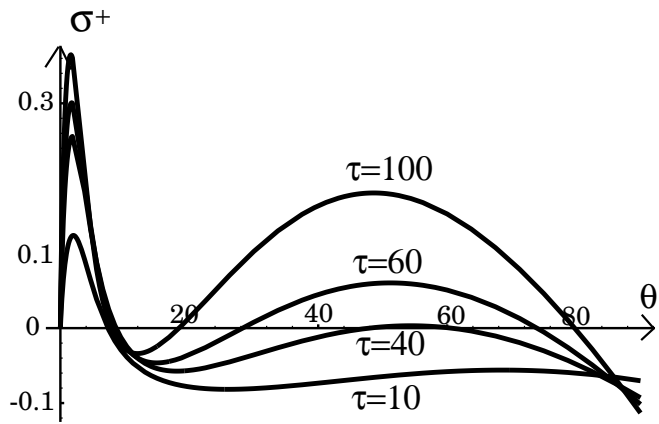}}
\caption{Growth rate $\sigma^{+}$, 
versus the perturbation angle $\theta>0$, for $P=15$,
$\epsilon=0.1$ and different values $\tau=10$, $40$, $60$, $100$ of
the rotation rate $\tau$.}
\end{figure}

\begin{figure}[htb]
\centerline{\includegraphics[width=8.6cm]{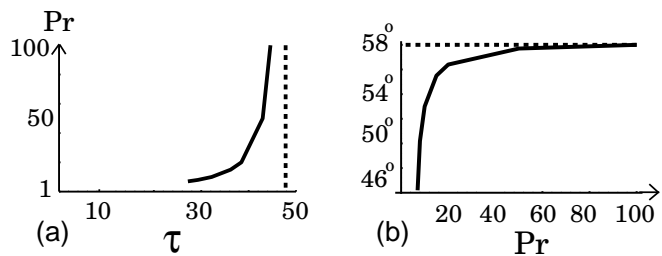}}
\caption{ K\"uppers-Lortz instability boundary in the 
($P_r, \tau$)-plane (a), and angle associated to the unstable perturbation
at the critical Taylor number versus the Prandtl number (b).}
\end{figure}

\section{Nature of the instability and nonlinear developments} 

We showed in Section 4 that in a rotating horizontal fluid layer
 with moderate Prandtl number, limited by top and bottom free-slip
boundaries, convective rolls are linearly
unstable with respect to perturbations in the form of rolls
making a small angle with that of the basic pattern. This instability
occurs even when the rotation rate is too low for the existence of the 
K\"uppers-Lortz instability.
It is related to the divergence of the growth rate
(\ref{eq:diverg0}) which, at finite Prandtl number, occurs  
when the direction of the wavevector of the
perturbation, approaches that of the basic rolls. We here chosed the associated
wavenumbers to be critical, but the effect survives whatever their values.

The above instability was obtained in an infinite domain. Its
persistence with lateral boundaries requires the
presence of a large number of rolls, and thus a convective cell with a large
aspect ratio $\mu^{-1}$. The mesh size in
Fourier space scaling like $\mu$, the minimum  angle $\theta$
between two wavevectors, behaves like $\mu^{1/2}$. Since for the
small-angle instability $\sigma\sim\epsilon^{4/3}$ and
$\theta\sim \epsilon^{2/3}$, it follows that $\sigma\sim \mu$,
a growth rate intermediate between that (of order unity) of 
a pure amplitude instability and a phase instability, for which
$\sigma$ scales like $\mu^2$.

As shown in \cite{ZS} and \cite{B-B} in the absence of rotation,
that of parallel rolls may also be unstable (for wavenumbers larger than
critical) to a skewed-varicose instability whose growth rate also
varies like the inverse aspect ratio $\mu$, a scaling resulting
from the strong magnitude of the mean flow in the case of free-slip boundary
conditions. This instability is however not captured by the present 
formalism since the roll distortions involved in this instability cannot
be represented within the class of perturbations (superposition of
two families of straight rolls), we have considered.

In order to investigate the relation between the small-angle and the
skewed-varicose instabilities, and to analyze their nonlinear
developments, a system of equations in the spirit of the Swift-Hohenberg 
equation, but coupling the leading vertical mode to the mean flow,
was derived by a systematic perturbation expansion near threshold
\cite{Ponty1}.  This
system which preserves the rotational invariance of the problem, 
generalizes equations obtained by Manneville \cite{Manneville} at 
finite Prandtl number in the absence of rotation. In a simplified
version where the non-local couplings are suppressed and only a few
representative nonlinear terms are retained, it is also
consistent with models used in \cite{Haken} and \cite{Haken2} for
rotating convection at infinite Prandtl number.
A similar model was considered in \cite{Xi}.
 
As discussed in \cite{Ponty1}, the phase equation derived in the context of
the generalized Swift Hohenberg equations, shows that
the skewed varicose instability occurring without rotation near
onset, becomes asymmetric with respect to the angle of the phase
perturbation, in the presence of rotation. This model also shows that both
the asymmetric skewed varicose and the small angle instabilities
lead, by means of reconnection to a progressive rotation
of the convective rolls in the
direction of the external rotation, an effect of the mean
flow which develops shear layers.

We are thus, led to conclude that the small-angle divergence 
of the K\"uppers-Lortz
instability growth rate pointed out in \cite{Knobloch},
results from the presence of a small-angle instability, 
which can be identified as an asymmetric skewed-varicose instability.
In contrast with the (symmetric) skewed varicose instability arising
without rotation, the asymetric one developing in presence of rotation
exists whatever value of the basic roll wavenumber.
Both instabilities are produced by the mean flow and 
deaseapear in the limit of infinite Prandtl numbers.

\bigskip
\noindent
{\bf Acknowlegments:}

\noindent
We are grateful to F.Busse for suggesting the relation with the skewed
varicose instability and to the anonymous referee for
pointing out reference.\cite{Xi} to our attention.

\end{document}